%% file: paper.tex
\documentclass[10pt, sigconf]{acmart}

\renewcommand\footnotetextcopyrightpermission[1]{} 
\setcopyright{none}

\include{macros}

\begin{document}
\title{\name: Fast Congestion Response using Explicit Incast Notifications for Datacenter Networks} 

\author{Hamidreza Almasi\qquad Hamed Rezaei\qquad Muhammad Usama Chaudhry\qquad Balajee Vamanan\\University of Illinois at Chicago}

\renewcommand{\shortauthors}{}

\begin{abstract}
\input{s-abstract}
\end{abstract}

\maketitle

\input{s-introduction}

\input{s-background}

\input{s-design}

\input{s-results}
\input{s-related}
\input{s-conclusion}

\bibliographystyle{ACM-Reference-Format}
\bibliography{references}

\end{document}

%% file: macros.tex
\usepackage{xspace}
\newcommand{\name}{{Pulser}\xspace}

\usepackage{enumitem}

\newcommand{\tabput}[3]{
\begin{table}
\begin{center}
{
#2
}
\end{center}
\caption{#3 \label{tab:#1}}
\vspace{-0.35in}
\end{table}
}

\newcommand{\figput}[4][1.0\linewidth]{
\begin{figure}[t]
\begin{minipage}{\linewidth}
\footnotesize 
\begin{center}
\includegraphics[width=#1]{figures/#2}
\end{center}
\vspace{-0.2in}
\caption{#4 \label{fig:#2}}
\end{minipage}
\end{figure}
}

\usepackage[linesnumbered,ruled,vlined]{algorithm2e} 

\usepackage{comment}

\usepackage{booktabs} 
\usepackage{array}

\makeatletter
\newcommand{\thickhline}{%
    \noalign {\ifnum 0=`}\fi \hrule height 1pt
    \futurelet \reserved@a \@xhline
}
\newcolumntype{"}{@{\hskip\tabcolsep\vrule width 1pt\hskip\tabcolsep}}
\makeatother

%% file: s-abstract.tex
Datacenter applications frequently cause incast congestion, which degrades both flow completion times of short flows and throughput of long flows. Without isolating incast, existing congestion control schemes (e.g., DCTCP) rely on existing ECN signal to react to general congestion, and they lose performance due to their slow, cautious, and inaccurate reaction to incast. We propose to isolate incast using Explicit Incast Notifications (EIN) that are generated by switches, similar to ECN. Our incast detection is fast and accurate. Further, we present our congestion control scheme, called \textit{\name}, which drastically backs off during incast based on EIN, but restores sending rate once incast ends. Our real experiments and ns-3 simulations show that \name outperforms prior schemes, DCTCP and ICTCP, in both flow completion times and throughput. 

%% file: s-introduction.tex
\section{Introduction} \label{sec:introduction}

Datacenters provide fast, curated access to vast amounts of Internet data. Today's datacenters host a mix of applications -- foreground applications perform distributed lookup in response to user queries and background applications perform data update and reorganization. While foreground applications predominantly generate short flows and the nature of distributed lookup implies that their performance is sensitive to higher percentiles (i.e., tail) of short-flow completion times~\cite{tail-cacm}, background applications generate long lasting flows and require high throughput. Therefore, today's datacenter networks optimize short-flow completion times and long-flow throughput. 

The key to optimizing both flow completion times (of short flows) and the throughput (of long flows) fundamentally lies in \textit{accurately} and \textit{quickly} responding to congestion. Traditional TCP uses packet loss to modulate its sending rate and relies on duplicate ACKs and timeouts to infer packet loss. Because packet loss is often a late indication of congestion, today's datacenter networks leverage some form of Active Queue Management (AQM) such as Explicit Congestion Notification (ECN), to quickly infer congestion. Current state-of-the-art datacenter networks use variants of DCTCP~\cite{dctcp}, which leverages ECN. ECN-enabled routers mark packets if their instantaneous queue length exceeds a predefined threshold and DCTCP senders modulates their sending rate proportional to the fraction of observed ECN marks in the ACK packets. 

While DCTCP senders respond to congestion faster than traditional TCP using early network feedback (i.e., ECN), DCTCP incurs packet drops when network queues buildup at a much faster \textit{rate} than DCTCP senders can respond; we show this phenomenon later in our results. Indeed, many foreground datacenter applications that, by design, perform distributed lookup for small data items spread across hundreds or thousands of servers, and, therefore, cause frequent \textit{incasts} (i.e., data from many input ports converges to one output port and cause rapid queue buildup). Today's incast-heavy applications (e.g., Web Search) and high-bandwidth network topologies (e.g., fat-trees with low over-subscription factors) imply that congestion often happens \textit{due to incasts} at the network edge, as reported by Google~\cite{google-dc} and Microsoft~\cite{microsoft-dc}. Because incast causes a rapid queue buildup in a short time, DCTCP's iterative, gradual window adaptation might not prevent buffer overflow. A more aggressive window adaptation algorithm or lower ECN threshold at the switch would cause throughput loss~\cite{ecn-multiqueue}.

In this paper, we make the case for isolating incast from other general cases of congestion. Because incast congestion is the common case, accurate detection and timely response to incast can significantly improve network performance, as our results show. Because detecting incasts at the end-hosts would require multiple roundtrips and would be significantly less efficient due to the short incast time scales, we argue for detecting incasts at switches, as opposed to detecting at end-hosts. We present a novel algorithm for detecting incasts in a short time interval by monitoring the gradient of queue length over small time windows. Similar to ECN, switches set an Explicit Incast Notification (EIN) mark upon detecting incasts. Switches detect incast per output port and mark packets traversing through those ports. 

We propose a DCTCP variant, called \textit{\name}, which leverages EIN for window adaptation. \name resets the congestion window to a small value upon observing EIN marked ACKs. While incasts last only for a short time and contribute to a small fraction of the overall network load, drastically resetting the congestion window only to ramp-up soon after would cause throughput loss. Therefore, \name restores the congestion window to its pre-incast value if subsequently received ACKs do not have EIN marks. The net effect is that \name has a braking phase when EIN marks are observed, which only lasts for a short time; after the incast episode, \name restores its pre-incast sending rate instead of a gradual increase. Fast and accurate incast detection is key to \name's design, and without such detection, \name (or DCTCP) would lose throughput. ICTCP~\cite{ictcp} addresses incast at the receiver without adding network support. Consequently, ICTCP's end-host detection is slow and \name outperforms ICTCP (see section \ref{sec:sims-results}). 

In summary, we make the following contributions: 
\begin{itemize}[noitemsep,topsep=0pt,parsep=0pt,partopsep=0pt]
\item We propose a combination of in-network and end-host mechanisms that specifically target incast congestion, which is common but not efficiently handled by existing proposals. 
\item We introduce a novel, gradient-based incast detection logic in switches, which is fast and accurate. 
\item We propose a congestion control scheme that uniquely leverages our incast detection to improve both short-flow completion times and long-flow throughput. 
\end{itemize}

Using a combination of real testbed and ns-3~\cite{ns3} simulations, we show that \name improves both $99^{th}$-percentile short-flow completion times and long-flow throughput:

\noindent \textbf{With simulations}, \name:
\begin{itemize}[noitemsep,topsep=0pt,parsep=0pt,partopsep=0pt]
\item achieves $10\%$ (1.12x) reduction in median and $50\%$ (2x) reduction in $99^{th}$ percentile FCT than DCTCP and ICTCP, on average, for loads greater than $20\%$. At higher loads, \name achieves up to $25\%$ and $70\%$ reduction in median and $99^{th}$ percentile FCT, respectively. 
\item achieves $20\%$ higher long-flow throughput than DCTCP and ICTCP, on average, for loads greater than $20\%$. \name achieves up to $50\%$ higher throughput at higher loads. 
\end{itemize}

\noindent \textbf{With real testbed}, \name:
\begin{itemize}[noitemsep,topsep=0pt,parsep=0pt,partopsep=0pt]
\item outperforms DCTCP by about $26\%$ in $99^{th}$ percentile flow completion times.
\item achieves about $25\%$ higher throughput than DCTCP.
\end{itemize}

The remainder of the paper is as follows. We start the motivation for our paper in section ~\ref{sec:background}, following by our design in section~\ref{sec:design}. Sections~\ref{sec:sims-results} and ~\ref{sec:results-real} present our experimental methodology and results. We discuss related work in section~\ref{sec:related} and conclude in section~\ref{sec:conclusion}.

%% file: s-background.tex
\section{Motivation} \label{sec:background}

DCTCP is a pioneering work that made a key insight that an accurate, proportional response to congestion using ECN could improve both flow completion times and throughput. DCTCP assembles 1-bit ECN marks at the end-host for a sequence of packets to infer accurate queue length at the bottleneck switch and uses the information to modulate the sending window~\cite{dctcp-analysis} accordingly. DCTCP performs well for long flows or when incast is somewhat mild. However, DCTCP's performs poorly with an incast-heavy traffic with many short flows. The queue size would increase rapidly during incast, and therefore, it is essential to \textit{drastically} slow down all senders in order to avoid packet loss. However, DCTCP's proportional response would require a few round-trips (RTTs), which is sub-optimal for incast.

Our at-scale ns-3 simulations capture this behavior. Figure~\ref{fig:dctcp} shows the evolution of a specific switch's queue length (red line) over time (X-axis). Figure~\ref{fig:dctcp} also shows DCTCP's reaction (green line), which is either 0 (DCTCP does not slow down) or 1 (DCTCP slows down due to ECN). We clearly see that even though incast starts at $time=10$, DCTCP does not react until $time=210$ when it is too late (i.e., DCTCP does not reduce its congestion window to the desired level until $time=210$). Please see section~\ref{sec:sims-results} for topology and workload details. While one could think of reducing ECN threshold to improve DCTCP's reaction to incasts, past papers have shown that smaller ECN thresholds detect incast spuriously and cause throughput loss~\cite{ecn-multiqueue, dctcp}. Later, we show that \name is able to react much faster than DCTCP (section~\ref{sec:sims-results}). 

\figput[0.85\linewidth]{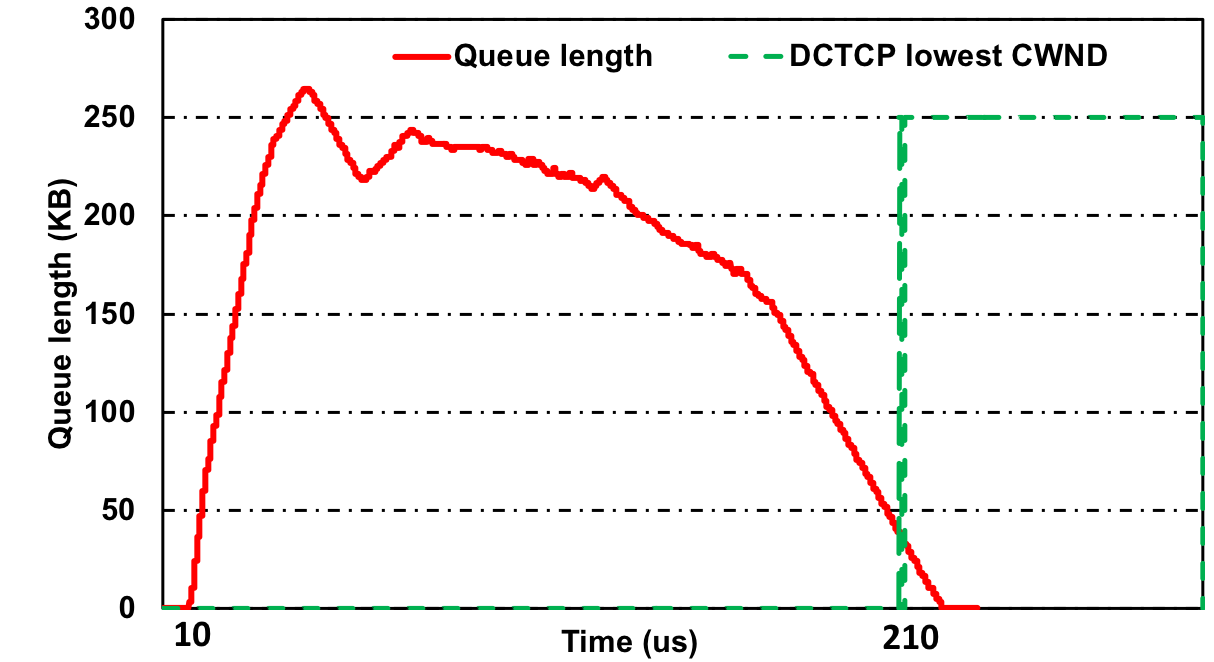}{}{Incast detection in DCTCP using ECN}

Though datacenter traffic is heavy tailed with a small fraction of long flows accounting for the majority of bytes transferred, the growing popularity of online services (e.g., Google Search, Facebook) implies that the fraction of short flows and the intensity of incast is bound to increase. While the performance of all end-to-end rate control schemes degrade as the fraction of short flows increase, we contend that an AQM scheme that is customized for incast detection and a congestion control algorithm that leverages the scheme could substantially improve performance over the current state-of-the-art.

%% file: s-design.tex
\section{\name} \label{sec:design}

\name consists of two parts: (1) fast and accurate incast detection and (2) end-to-end congestion control that leverages incast detection. We describe our novel incast detection in section~\ref{sec:design-ein} and our congestion control in section~\ref{sec:design-cc}. 

\subsection{Explicit Incast Notifications (EIN)} \label{sec:design-ein}

\begin{algorithm}
\DontPrintSemicolon
\SetAlgoLined
\KwResult{Set or Reset EIN}
\SetKwInOut{Input}{Input}\SetKwInOut{Output}{Output}
\Input{$Qlen$}
\Output{EIN}
\BlankLine

\For{Each packet ``P'' at dequeue}
{
    $Gradient = (Qlen - Qlen_{prev}) / (T - T_{prev})$\;
    $Qlen_{prev} = Qlen$\;
    $T_{prev} = T$\;
    Store $Gradient$ in a sliding window\;
    Calculate $Average~Gradient$ for ``N'' samples\;
    \eIf{$Average~Gradient > EIN_{threshold}$}
    {
        Set EIN\;
        $EIN_{prev} = 1$\;
    }
    {
        \eIf{$EIN_{prev} == 1$}
        {
            \If{$Qlen > HighWaterMark$}
            {
                Set EIN\;
            }
        }
        {
            Reset EIN\;
            $EIN_{prev} = 0$\;
        }
    }
    
}
\caption{EIN generation at switches}\label{alg:incast-detection}
\end{algorithm}
\bigskip 

During incast, data from multiple input ports (e.g., > 8) gets forwarded to the same output port within a switch, which cause a steep increase in the output port's queue length. Therefore, our incast detection logic uses the gradient of queue length as opposed to the queue length itself. 

Similar to most ECN implementations, we perform incast detection at the dequeue side. At a high level, we calculate the gradient during each dequeue event w.r.t to previous event (i.e., slope between two consecutive packets that were dequeued). We maintain a sliding window of past $N$ samples of gradient, where $N$ is configurable. If the average gradient is more than a configurable threshold, then we mark outgoing packets by setting the new Explicit Incast Notification (EIN) bit. EIN requires one additional bit in the IP header (similar to CE bit for ECN), which is set by the switches and another bit in the TCP header for notify senders (similar to ECE for ECN). 

Algorithm~\ref{alg:incast-detection} shows our complete algorithm. In our implementation, we also set EIN when the current queue length exceeds a configurable $HighWaterMark$, which serves as hysteresis (see lines 11-18). We set $HighWaterMark$ to be higher than ECN threshold to avoid throughput loss. Our incast detection has two main parameters, $N$ and $EIN_{threshold}$. We empirically found that using $N=50$ and $EIN_{threshold}=0.25 \times Line~Rate$ provides optimal performance. Intuitively, our parameter settings mean that if the queue is building at the rate of $0.25 \times Line~Rate$ for the past $50$ samples on average, then we detect an incast episode. Consequently, we detect incast if either the queue builds up steeply in a very short time window or if the queue consistently builds up over a long time window. In either case, it is is desirable to react strongly by setting the EIN bit to avoid buffer overflow. We also performed an exhaustive sensitivity study but do not show due to space constraints. Our incast detection is fast and accurate, as we show in section~\ref{sec:sims-results}. 

\subsection{Congestion control} \label{sec:design-cc}

We design \name's congestion control by leveraging EIN. If a \name sender gets a packet with EIN mark, the sender reduces its congestion window to a configurable, \textit{safe} value after saving the current congestion window. Such a drastic response to incast congestion would likely ease congestion. Once incast finishes, the sender would stop receiving EIN marks. If the sender doesn't not observe any EIN marked packets for the current batch of packets, then the sender restores the window to its previous \textit{saved} value. Equations~\ref{eq:cc-set} and~\ref{eq:cc-reset} show how we modify the congestion window at the beginning and end of an incast episode, which we infer via EIN marks. 

\begin{equation}\label{eq:cc-set}
\begin{split}
  cwnd_{prev} \gets cwnd \\
  cwnd \gets cwnd_{safe}
\end{split}
\end{equation}

\begin{equation}\label{eq:cc-reset}
  cwnd \gets cwnd_{prev}
\end{equation}

We empirically found out that setting $cwnd_{safe}=4 \times MSS$ provides optimal performance. We did a sensitivity study but do not show due to lack of space. As you can see, our congestion control is only a handful lines of code change over existing DCTCP implementation and is deployment friendly. 

%% file: s-results.tex
\section{Simulation Methodology}\label{sec:sims-method}

We use ns-3 \cite{ns3} to simulate a leaf-spine datacenter topology, which is commonly used in today's datacenters~\cite{CONGA}. In our topology, the fabric interconnects $400$ servers using $20$ leaf switches with each leaf switch connecting to 20 servers. The leaf switches are connected to 10 spines, resulting in an over-subscription factor of $2$. The servers and switches are connected by $10~Gbps$ links with an unloaded link delay of $10~{\mu}s$; the unloaded Round-Trip Time (RTT) for the longest path (i.e., $4$ hops) is 80 $\mu$s. 

We model our workloads based on reported results~\cite{benson}, with a mix of short and long flows. Flow arrivals follow a Poisson distribution and the source and destination for each flow is chosen uniformly randomly. Our short flows' sizes are randomly chosen from $8~KB$ to $32~KB$ and we set long flow sizes to $1~MB$. As as typical, long flows contribute to $30~\%$ of the overall network load, which we vary in our experiments~\cite{pfabric}. We also model incast traffic as per~\cite{incast2009safe}. The flows and their destinations are chosen randomly and are varied during the experiment. Our default incast degree is $24$ but vary it in our sensitivity analysis~\ref{sec:sims-sens}. 

We compare four schemes: DCTCP, ICTCP, \name, and \textit{Ideal}. Our DCTCP and ICTCP implementations use their recommended parameter settings (e.g, ECN threshold) and our results match their reported numbers. We implemented {\name} on top of DCTCP \cite{dctcp}. We implemented algorithm~\ref{alg:incast-detection} in switches and our congestion control in end-hosts. We set: $cwnd_{safe} = 4 \times MSS$, $EIN_{threshold} = 0.25 \times Line~Rate$, and $N = 50$ as default, after sensitivity studies (not shown due to lack of space). We also implemented an \textit{Ideal} congestion control scheme where senders have \textit{oracular} global knowledge and send at optimal sending rate. While the Ideal scheme is not practical, we show its results to set reasonable upper bounds on performance. 

\section{Simulation Results} \label{sec:sims-results}

We summarize our evaluation of \name as follows: 
\begin{itemize}
    \item \textbf{Flow Completion Time (FCT)}: We compare the median and $99^{th}$ percentile short-flow completion times of \name with DCTCP, ICTCP, and Ideal. \name achieves $10\%$ (1.12x) reduction in median and $50\%$ (2x) reduction in $99^{th}$ percentile FCT than DCTCP and ICTCP, on average, for loads greater than $20\%$. At higher loads, \name achieves up to $25\%$ and $70\%$ reduction in median and $99^{th}$ percentile FCT, respectively. 
    \item \textbf{Throughput}: We compare the long-flow throughput of \name with DCTCP, ICTCP, and Ideal. \name achieves $20\%$ higher long-flow throughput than DCTCP and ICTCP, on average, for loads greater than $20\%$. \name achieves up to $50\%$ higher throughput at higher loads. 
    \item \textbf{Queue length analysis}: We analyzed how the queues buildup in \name and DCTCP. \name reduces queue lengths drastically (by up to 2x) compared to DCTCP. 
    \item \textbf{Sensitivity to incast}: \name's improvements increase with increasing incast degree and is robust across a range of typical incast degrees. 
\end{itemize}

We provide a more exhaustive analysis below. 

\subsection{Flow Completion Time}

\figput[0.75\linewidth]{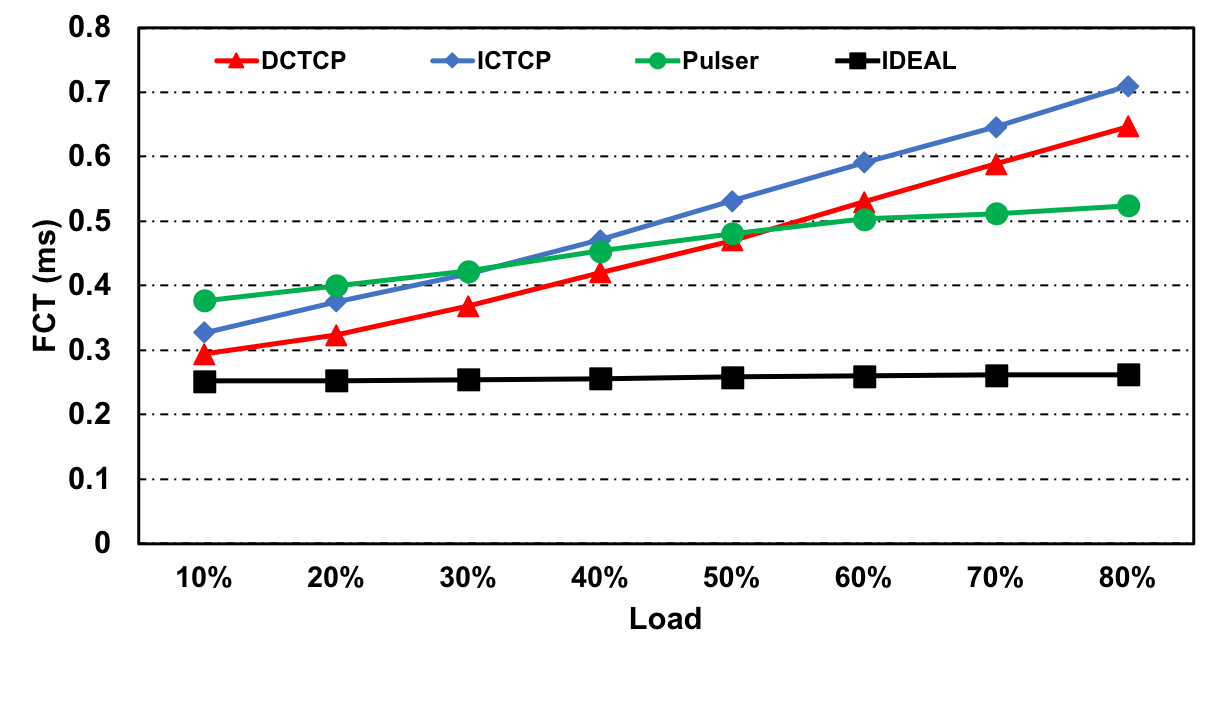}{}{Median flow completion time}
\figput[0.75\linewidth]{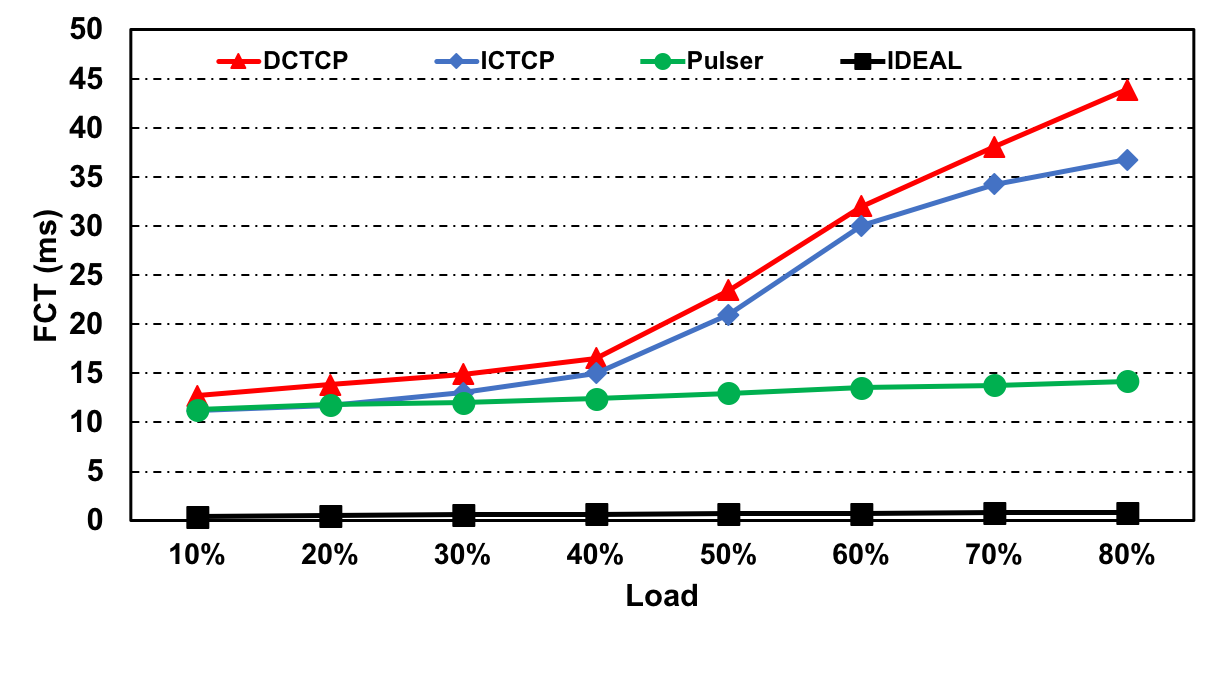}{}{$99^{th}$ \%-ile flow completion times}

Figure ~\ref{fig:50th} and figure ~\ref{fig:99th} compare the median and tail ($99^{th}$ percentile) flow completion times of DCTCP, ICTCP, \name, and Ideal. We show flow completion times along Y-axis versus network load on X-axis. As load increases, all schemes incur more queuing and their FCTs degrade. While \name achieves reduction in both median and tail FCT, \name's achieves better reduction in tail flow completion times than in median flow completion times. Because datacenter applications are more sensitive to tail FCT than median, \name's makes the right trade-off. 

Compared to DCTCP, \name reduces tail flow completion time of about 51\% for loads greater than 40\% (typical operating point of most datacenters). Compared to ICTCP, \name reduces flow completion time by about 46\% at higher loads. Because incast congestion is not an issue at lower loads, \name does not significantly outperform at lower loads. \textit{Ideal} method outperforms all other schemes, which shows that there is significant improvement to be achieved. 

\subsection{Throughput}

\figput[0.75\linewidth]{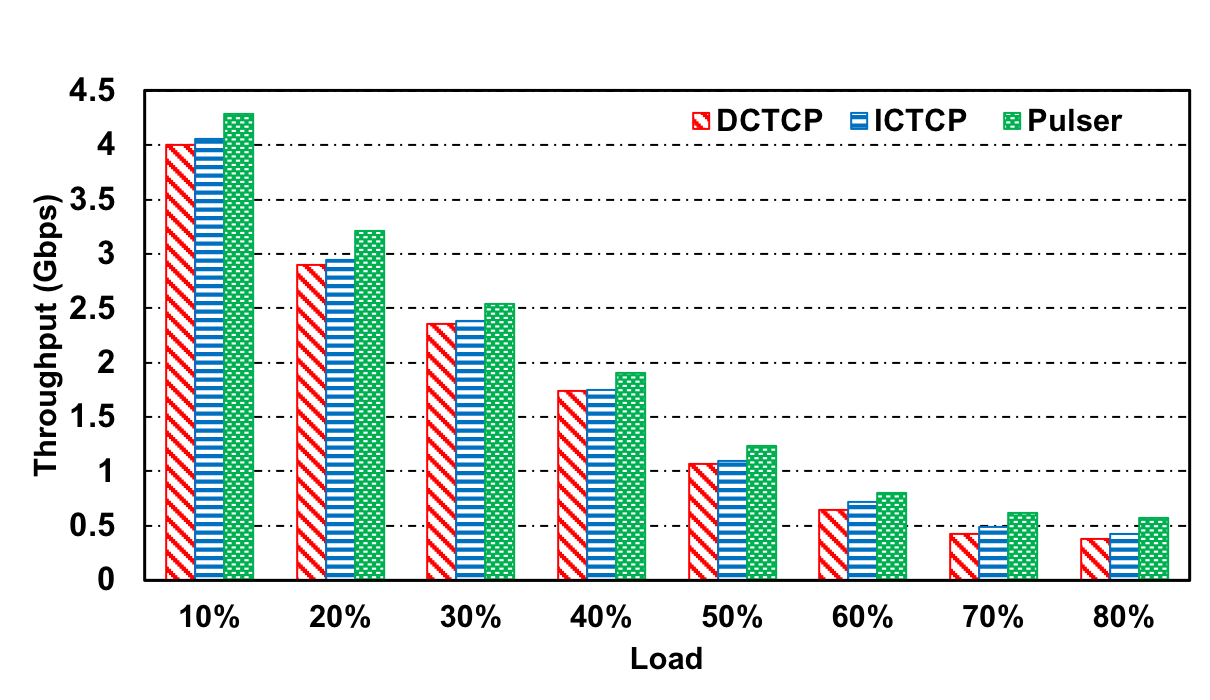}{}{Throughput comparisons}

In this section we compare {\name}'s throughput with ICTCP, DCTCP and ideal scheme. Long flow throughput suffers for all schemes at higher loads due to increased (incast) congestion at higher loads.  As we can see from figure~\ref{fig:throughput}, \name achieves higher throughput across all loads: First, \name reduces the number of packet drops of those background flows that share links that experience high incast congestion, as compared to other schemes. Second, when incast is finished, \name uses the last congestion window before incast as the new congestion window, without resorting to gradual window increase (e.g., slow start). \name's ON/OFF window modulation helps senders to restore their pre-incast sending rate pretty quickly.\name achieves 16\% and 22\% higher throughput compared to ICTCP and DCTCP respectively (in 40\% load and beyond).

\subsection{Queue length} 

\figput[0.75\linewidth]{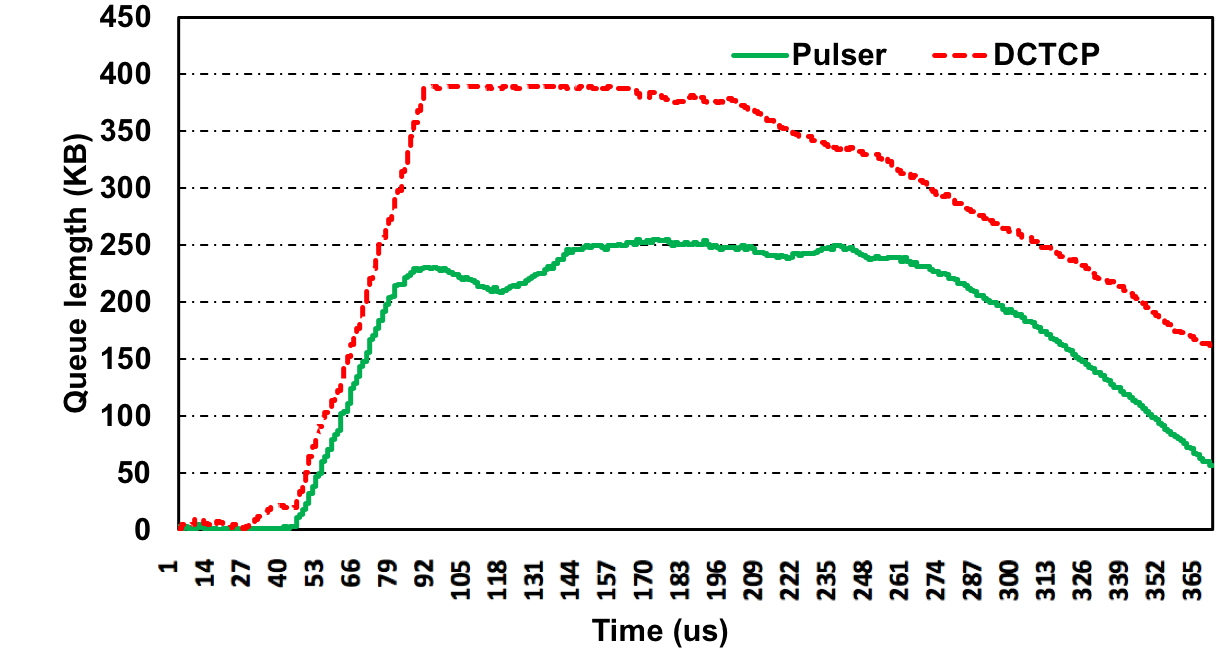}{}{Queue length over time}
\figput[0.75\linewidth]{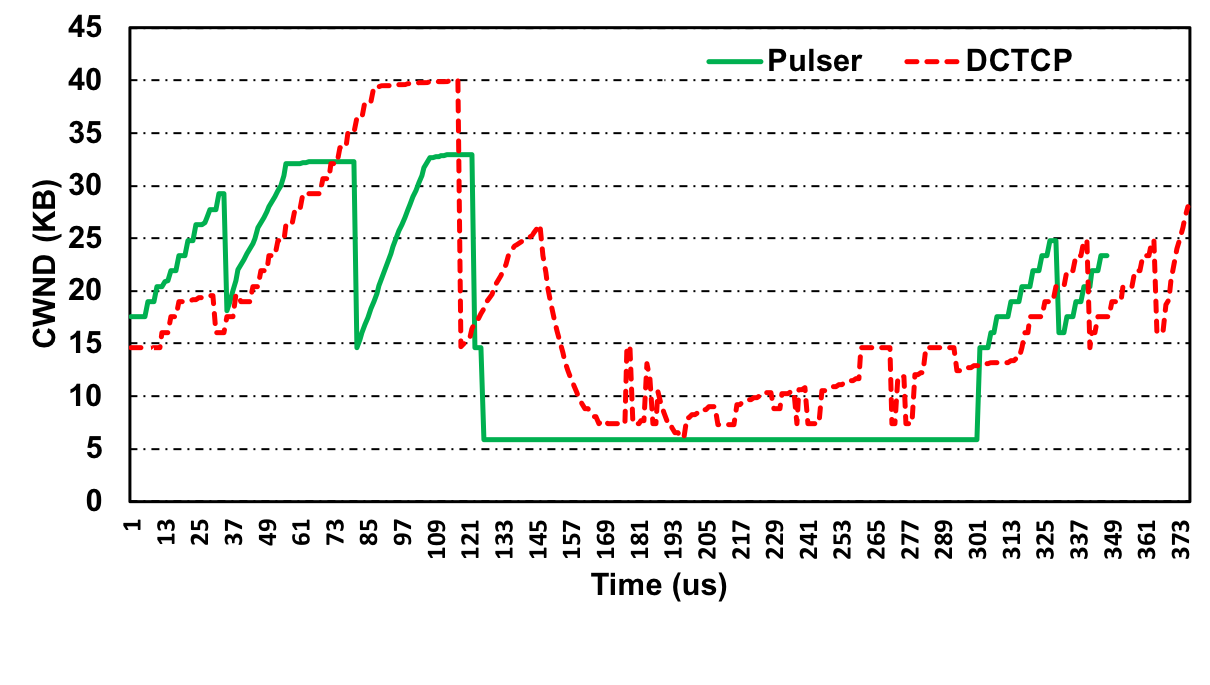}{}{Congestion window at a long flow sender}

In this section, we analyze the queuing behavior of \name and relate it to \name's congestion control (i.e., evolution of congestion window over time). For this experiment, we run our workload with $60\%$ load. Figure~\ref{fig:qlen} shows the queue length at an aggregator switch's output port (Y-axis) over time (X-axis). We analyze DCTCP (red) vs. \name (green). We see that, \name reduces the queue buildup by as much as $50\%$ (2x). 

To connect \name's queuing behavior to our congestion control, we compare the congestion window evolution versus time (at the sender) for DCTCP and \name in figure~\ref{fig:cwnd}. At $time = 120{\mu}s$, incast starts. While DCTCP gradually reduces the congestion window and oscillates around due to the absence of a precise signal that indicates incast, \name leverages a more precise EIN to backup almost instantly. When the incast finishes at $time = 300{\mu}s$, \name instantly recovers. By instantly backing off, the \name's long-flow sender minimizes queuing delay, which helps short flows. By restoring its previous sending rate after incast, \name sender achieves better throughput. 

\subsection{Sensitivity to incast degree}\label{sec:sims-sens}
We analyze the sensitivity of our results to different incast degrees. For this study, we compare \name's tail flow completion time to those of DCTCP and ICTCP for varying incast degrees. We vary incast degree as 24 (default), 32, and 40. Figure~\ref{fig:sensitivity} shows the $99^{th}$ percentile flow completion times for varying incast degrees, normalized to our default case (i.e., incast degree of 24). 

\figput[0.75\linewidth]{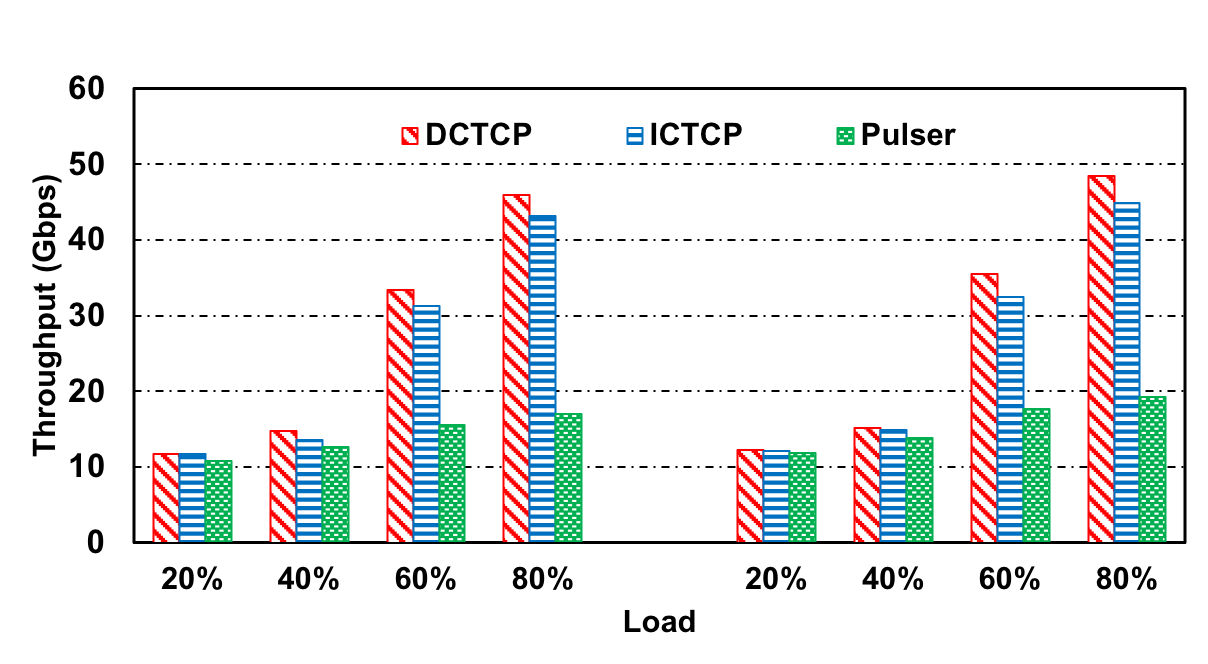}{}{Sensitivity of $99^{th}$ percentile flow completion times to incast degree}

As expected for both incast degrees, all methods experience increasing tail flow completion time with load increments. In both cases, \name outperforms DCTCP and ICTCP with a substantial margin of at least 2X for 60\% and 80\% loads. Lower loads do not suffer from high incast congestion, and, therefore, there is limited opportunity for improvement. Similar to higher loads, higher incast degree provide more opportunity for \name. Nevertheless, \name's relative performance improvement remains robust across varying loads and incast degrees. 

\section{Real implementation} \label{sec:results-real}

Our real testbed consists of three Dell 7040 \textit{Optiplex} servers with 32 GB of memory, Intel Quad core processors (3.4 GHz i7) and 1 Gbps NICs. Two servers act as clients and generate traffic to the third server, which acts as an aggregator (leaf server). Because EIN requires switch support, we use another server with two network interfaces as a software switch (kernel version 4.4.0). The two client servers are connected by a physical Netgear \textit{Prosafe} switch to our software switch, which connects to the aggregator. Further, to generate a realistic incast scenario with only two servers, we place 8 VMs in each of the two client servers; the VMs run Ubuntu 12.04 LTS (kernel version 3.2.18) with 2GB of memory. We rate limit the client VM's NICs to 50 Mbps.The two client machines each generate $50 \times 8 = 400 Mbps$ of traffic to the physical switch, which connects to the software switch over a 1 Gbps link (i.e., there is no bottleneck). However, the link between the software switch and the aggregator is rate limited to 50 Mbps, creating a realistic incast (i.e., there is 800 Mbps of incoming traffic into the software switch but the outgoing port is only 50 Mbps, which creates a realistic incast degree of 16). We use \textit{iperf3} to generate traffic. We generate a background 40 MB long flow from one of the client VMs. The other 15 client VMs generate synchronous bursts of short 100KB flows, with random jitter. We run the experiment for 80 minutes and measure the flow completion times of short flows and throughput of long flows. 

Table~\ref{tab:table-real} shows the flow completion times -- both average (not median) and $99^{th}$ percentile -- and throughput comparison between DCTCP and \name in our real testbed. Our real testbed is smaller in scale, and, therefore, the intensity of incast and the corresponding tail effects are somewhat less pronounced in our real testbed than in our at-scale simulations. Nevertheless, \name outperforms DCTCP by about $20\%$ and $26\%$ in average and $99^{th}$ percentile flow completion times, respectively. Similarly, \name achieves about $25\%$ higher throughput than DCTCP. While we do not have access to a datacenter-scale testbed, our substantial performance gains in the small testbed shows the potential of \name in a more realistic setting. 

\tabput{table-real}{
\small
\begin{tabular}{|c"c|c|}
 \thickhline
 \textbf{Metric} & \textbf{DCTCP} & \textbf{Pulser}\\ 
 \thickhline
 \textit{Avg. flow completion time} (s) & $1.99$ & $1.59$\\ 
 \hline
 \textit{$99^{th}$ percentile flow completion time} (s) & $13.32$ & $9.85$\\ 
 \thickhline
 \textit{Throughput} (Mbps) & $28$ & $35$\\ 
 \thickhline
\end{tabular}
}{Real implementation results}

%% file: s-related.tex
\section{Related work} \label{sec:related}

While Internet Congestion control is a well-studied research area, datacenter congestion control continues to garner interest in the networking community and there are a number of recent papers on datacenter congestion control. We have discussed DCTCP and ICTCP in earlier sections. We will summarize other related work in this area. 

Rate Control Protocol (RCP)~\cite{rcp} is an alternative to window-based TCP protocols in which switches directly inform the senders of their fair share sending rate by observing the rates of all intervening flows. But, RCP does not isolate incast, and requires switch support, which is not available today. 
Similar to \name, TIMELY~\cite{timely} uses a gradient-based approach. However, unlike \name, TIMELY is RTT-based, its detection logic is not customized for incast, and it is implemented at end-hosts. Therefore, TIMELY's detection is unlikely to be as fast and as accurate as our approach. 
DCQCN~\cite{dcqcn} leverages ECN for RDMA and performs rate-based congestion control. Our incast detection and congestion control ideas are complimentary to DCQCN and they would likely improve DCQCN's incast performance. 
QCN~\cite{qcn} provides congestion control based on network feedback (similar to DCTCP/ECN) but operated at the Ethernet layer and doesn't isolate incast. 
NumFabric~\cite{numfabric} provides other more flexible bandwidth allocations other than TCP's fair share. 
ExpressPass~\cite{expresspass} and NDP~\cite{ndp} provide receiver-driven congestion control; \name, in contrast, is switch-driven and isolates incast congestion from other forms of congestion (e.g., congestion in network caused by flow collisions). 
A number of proposals~\cite{d2tcp,pdq,detail,pfabric,karuna, ups, phost} focus on flow scheduling and prioritize critical flows (e.g., short flows) whereas our main focus is on incast congestion control. Similarly, other load balancing proposals~\cite{MPTCPSIGCOMM11,presto16,CONGA} are complimentary to \name. 

%% file: s-conclusion.tex
\section{Conclusion} \label{sec:conclusion}
Incast congestion is a dominant form of congestion in datacenter networks. Prior approaches do not isolate and detect incast in the network, and, therefore, existing end-host congestion control could not aggressively respond to incast without losing throughput. We proposed Explicit Incast Notification (EIN), a gradient-based incast detection at network switches, which is both fast and accurate. Leveraging EIN, we introduced our congestion control scheme, called \name, which quickly backs off during incast for short time intervals without hurting latency and ramps up soon after without losing throughput. Using simulations and a real implementation, we showed that \name outperforms DCTCP and ICTCP. As data and Internet traffic continue to grow exponentially, incast is likely to become even more dominant in datacenters, requiring an incast-specific AQM such as EIN and a congestion-control schemes such as \name.